\documentclass[structabstract]{aa}  
\usepackage{graphicx}
\usepackage{txfonts}

\usepackage{natbib}
\bibpunct[, ]{(}{)}{;}{a}{}{,}
\newcommand{\farc}{\hbox{$.\!\!^{\prime\prime}$}} % Fractions of arcseconds

\newcommand{\hb}{H$\beta$} 
\newcommand{\ha}{H$\alpha$} 
\newcommand{\oi}{[\ion{O}{i}]} 
\newcommand{\sii}{[\ion{S}{ii}]} 
\newcommand{\oii}{[\ion{O}{ii}]} 
\newcommand{\oiii}{[\ion{O}{iii}]}
\newcommand{\nii}{[\ion{N}{ii}]}

\begin{document}
\title{The metal-enriched host of an energetic $\gamma$-ray burst at $z\approx 1.6$ \thanks{Based on observations made with telescopes at the European Southern Observatory at LaSilla/Paranal, Chile under program 087.B-0737(C).}}

\titlerunning{Metallicity of the host of GRB 080605 at $z\sim 1.6$}

%078.D-0416, 177.A-0591,
%\subtitle{Physical conditions in the ISM of a vigorously star-forming, metal-rich GRB host at $z \sim 1.6$}

\author{T. Kr\"uhler\inst{1}, J.~P.~U. Fynbo\inst{1}, S. Geier\inst{1}, J. Hjorth\inst{1}, D. Malesani\inst{1}, B. Milvang-Jensen\inst{1}, A.~J.~Levan\inst{2}, M. Sparre\inst{1}, D.~J. Watson\inst{1}, and T. Zafar\inst{1} 
          }

\authorrunning{T. Kr\"uhler et al.}

\institute{
	Dark Cosmology Centre, Niels Bohr Institute, University of Copenhagen, Juliane Maries Vej 30, 2100 Copenhagen, Denmark. \and
	Department of Physics, University of Warwick, Coventry CV4 7AL, UK.
           }

          % School of Physics, University College Dublin, Dublin 4, Ireland
%            \and
%           European Southern Observatory, 85748 Garching, Germany.
%            }

\date{} %{Received September 15, 1996; accepted March 16, 1997}

% \abstract{}{}{}{}{} 
% 5 {} token are mandatory
 
\abstract
  {The star-forming nature of long $\gamma$-ray burst (GRB) host galaxies provides invaluable constraints on the progenitors of GRBs and might open a short-cut to the characteristics of typical star-forming galaxies throughout the history of the Universe. Due to the absence of near-infrared (NIR) spectroscopy, however, detailed investigations, specifically a determination of the gas-phase metallicity of gamma-ray burst hosts, was largely limited to redshifts $z < 1$ to date. }
{We observed the galaxy hosting GRB~080605 at $z=1.64$ using optical/NIR spectroscopy and {high-resolution HST/WFC3 imaging} in the rest-frame wavelength range between $1150$ and $8700~\AA$. These data allow us to study a $z > 1$ GRB host in unprecedented detail and investigate the relation between GRB hosts and field galaxies.}
{We avail of VLT/X-shooter optical/NIR spectroscopy to measure the metallicity, electron density, star-formation rate (SFR), and reddening of the host of GRB~080605. Specifically, we use different strong-line diagnostics to robustly measure the gas-phase metallicity within the interstellar medium (ISM) for the first time based on \nii~at this redshift.}
{The host of the energetic ($E_{\gamma,\rm{iso}} \sim 2\times 10^{53}$~erg) GRB~080605 at $z \sim 1.64$ is a {morphologically complex}, vigorously star-forming galaxy with an  H$\alpha$-derived SFR of $31^{+12}_{-6}~M_{\sun}\,\rm{yr}^{-1}$. {Its ISM is significantly enriched with metals. Specifically, \nii/H$\alpha = 0.14\pm0.02$ which yields an oxygen abundance $12 + \log(\rm{O}/\rm{H})$ between 8.3 and 8.6 depending on the adopted strong-line calibrator}. This corresponds to values in the range of $0.4-0.8\, Z_{\sun}$. For its measured stellar mass ($M_* = 8.0^{+1.3}_{-1.6}\times10^{9} M_{\sun}$) and SFR this value is consistent with the fundamental metallicity relation defined by star-forming field galaxies. The absence of strong Ly$\alpha$ emission constrains the escape fraction of resonantly-scattered Ly$\alpha$ photons to $f_{\rm{esc}} \lesssim 0.08$.}
{Our observations provide a detailed picture of the conditions in the ISM of a highly star-forming galaxy with irregular morphology at $z\sim1.6$. They include the first robust metallicity measurement based on \nii~for a GRB host at $z > 1$ and {directly illustrate that GRB hosts are not necessarily metal-poor, both on absolute scales as well as relative to their stellar mass and SFR. GRB hosts could thus be fair tracers of the population of ordinary star-forming galaxies at high redshift.}}

\keywords{Gamma-ray burst: general, individual: GRB~080605, ISM: abundances, Galaxies: star formation, high-redshift}

\maketitle
%
%________________________________________________________________

\section{Introduction}

The violent stellar explosion that gives rise to long $\gamma$-ray bursts \citep[see e.g.,][for reviews]{2004RvMP...76.1143P, 2009ARA&A..47..567G} and their multi-wavelength afterglows has been firmly related to broad-line supernovae (SNe) of type Ic, and hence star-formation (SF), via the core-collapse of massive stars \citep[e.g.,][]{1998Natur.395..670G, 2003Natur.423..847H, 2003ApJ...591L..17S, 2004ApJ...609L...5M, 2006Natur.442.1011P, 2006Natur.442.1008C}. The GRB's high-energy signature is very luminous, and unaffected by dust and therefore pin-points regions of star-formation irrespective of galaxy brightness, dust obscuration and redshift. GRB-selected galaxies hence provide a sample of high-redshift, star-forming galaxies that is fully complementary to conventional survey studies.

The luminous afterglows furthermore facilitate redshift measurements, and detailed investigation about the chemical composition \citep[e.g.,][]{2003ApJ...585..638S, 2006ApJ...648...95P, 2009ApJ...691L..27P, 2010A&A...513A..42D} and the dust properties of the host \citep[e.g.,][]{2001ApJ...549L.209G, 2006ApJ...641..993K, 2007MNRAS.377..273S, 2010MNRAS.401.2773S, 2011arXiv1102.1469Z}. {GRB hosts can hence be targeted with a known redshift, position and information about the galaxy's interstellar medium (ISM) at hand, providing an independent diagnostic of galaxy evolution and star-formation.}

{Notably at the highest redshifts \citep{2009ApJ...693.1610G, 2009Natur.461.1254T, 2009Natur.461.1258S, 0429BCucc}, GRBs allow us to set observational constraints on the history of star-formation \citep[e.g.,][]{2009ApJ...705L.104K, 2012ApJ...744...95R, 2012arXiv1202.1225E}, the galaxy luminosity function \citep{2012arXiv1201.6074T, 2012arXiv1201.6383B} as well as on the nature of young and star-forming galaxies \citep[e.g.,][]{2004A&A...425..913C, 2009ApJ...691..152C, 2010arXiv1010.1783W} beyond the detection limit of state-of-the-art surveys.}

{To represent a robust tool for cosmology and probe of star-formation, the physical conditions that lead to the formation of the GRB progenitor must be understood. As direct observations of GRB progenitors akin to those of some SNe remain impossible due to the cosmological distances, afterglow sight-line \citep[e.g.,][]{2009ApJS..185..526F}, spatially-resolved \citep[e.g.,][]{2008A&A...490...45C, 2008ApJ...676.1151T, 2011ApJ...739...23L} or galaxy-integrated measurements \citep[e.g.,][]{2009AIPC.1133..269G, 2012MNRAS.419.3039C} provide the most constraining information on the kind of galactic environments GRBs occur in.}

However, the properties of an unbiased sample of long GRBs hosts are still largely unknown, and selection effects due to optically-dark bursts \citep{1998ApJ...493L..27G, 2001A&A...369..373F, 2009AJ....138.1690P} arguably play a crucial role \citep[e.g.,][]{2011arXiv1108.0674K, 2011AAS...21710802P}. Consequently, the conditions for the formation of GRBs, the relation between GRB hosts and field galaxies and the extent to which GRBs trace the cosmic SFR remain highly debated \citep[e.g.,][]{2005MNRAS.362..245J, 2006Natur.441..463F,  2009ApJ...702..377K, 2011arXiv1105.1378C, 2011ApJ...735L...8K}. 

{Local galaxies hosting long GRBs tend to be of low stellar mass and metal content with respect to SDSS galaxies \citep{2010AJ....139..694L} as well as the hosts of core-collapse SNe \citep{2006Natur.441..463F, 2008AJ....135.1136M}, which has been interpreted as support for a limited chemical evolution of the GRB host - seemingly in line with metallicity constraints on the GRB progenitor from theoretical calculations based on the collapsar model \citep{1993ApJ...405..273W, 1999ApJ...524..262M}. These properties, however, are not indicative per-se of GRBs preferring low-metallicity environments, but instead could be the result of low-mass, low-metallicity galaxies dominating the local star-formation rate \citep{2011MNRAS.414.1263M}. In fact, at higher redshifts \citep[see e.g.,][]{2006ApJ...647..471L, 2007ApJ...660..504B} and in \textit{Swift} GRB host samples with a better controlled selection a population of red, luminous, high-mass hosts emerges \citep{Rossi2011, 2011ApJ...736L..36H, 2011ApJ...727L..53C, 2011arXiv1109.3167S}.}

A fundamental characteristic of (GRB-selected) galaxies is their gas-phase metallicity, and in particular whether they follow the relation between stellar mass ($M_*$), metallicity ($Z$) and SFR defined by local field galaxies \citep{2010MNRAS.408.2115M, 2010A&A...521L..53L}. However, observational access to the metallicity of GRB hosts remained largely elusive, and robust constraints are only available up to $z\sim1$ \citep{2009ApJ...691..182S, 2010AJ....139..694L}. This is largely due to the absence of efficient NIR spectrographs, as important tracers of metallicity {(such as \nii~($\lambda$6584) and H$\alpha$)}, are redshifted into the NIR wavelength regime above $z\sim0.5$. 

Here we present optical/NIR observations of the galaxy hosting GRB 080605  at $z = 1.64$ obtained with the X-shooter spectrograph at the Very Large Telescope (VLT), and NIR imaging {with HST/WFC3 and} LIRIS mounted at the William Herschel Telescope (WHT). The spectroscopic observations probe the rest-frame wavelength range between $1150$ and $8700\,\AA$~and reveal a wealth of emission lines including H$\beta$, \oiii, H$\alpha$ and \nii~($\lambda$6584). %The detection of \nii($\lambda$6584) allows us to measure the physical characteristics such as metallicity for the first time robustly for a GRB host at $z > 1$.

GRB 080605 \citep{2008GCN..7828....1S} was initially detected by the \textit{Swift} satellite \citep{2004ApJ...611.1005G}, and its optical/NIR afterglow was readily identified \citep{2008GCN..7829....1K, 2008GCN..7834....1C}. Spectroscopy of the afterglow was obtained with FORS2 at the VLT which yields a redshift of $z=1.6403$ \citep{2008GCN..7832....1J, 2009ApJS..185..526F}. The optical/NIR afterglow is characterized by the presence of significant amounts of dust with $A_V \sim 0.5$~mag \citep{2011A&A...526A..30G, 2011arXiv1102.1469Z}, including evidence of a 2175~\AA~feature \citep{2011Zafar}. The 2175~\AA~dust bump is a common characteristic observed along sight-lines through the Milky-Way. It becomes weaker in the Large Magellanic Cloud, and is absent from most sight-lines through the Small Magellanic Cloud. It is only rarely observed towards high-redshift environments such as quasars or absorbing systems, but common along sight-lines to highly extinguished afterglows \citep[e.g.][]{2008ApJ...685..376K, 2009ApJ...697.1725E, 2011arXiv1102.1469Z, 2010arXiv1009.0004P}. The carrier of the bump is currently not fully understood with graphite and polycyclic aromatic hydrocarbons being primary candidates \citep[see e.g.,][for a review]{2003ARA&A..41..241D}. %Galaxies hosting sight-lines with such a dust feature are thus of special interest, and a photometric study already revealed them to be more massive than the bulk of unextinguished afterglow hosts \citep{2011arXiv1108.0674K}. % Here we present the first detailed spectroscopic observations of such a host galaxy.

We adopt the concordance ($\Omega_M=0.27$, $\Omega_{\Lambda}=0.73$, $H_0=71~\rm{km}\,s^{-1}\,\rm{Mpc}^{-1}$) $\Lambda$CDM cosmology. All errors are given at $1\sigma$ confidence levels. All magnitudes are given in the AB system and {are corrected for the Galactic reddening of $E_{B-V}=0.137$~mag \citep{1998ApJ...500..525S}. The solar oxygen abundance is assumed to be $12+\log(\rm{O}/\rm{H}) = 8.69$ \citep{2009ARA&A..47..481A} throughout this work.} Wavelengths are given in vacuum and the redshifts in the heliocentric system.

\section{Observations and data reduction}

\subsection{Space-based imaging}

\begin{figure}
\centering
\includegraphics[width=0.99\columnwidth]{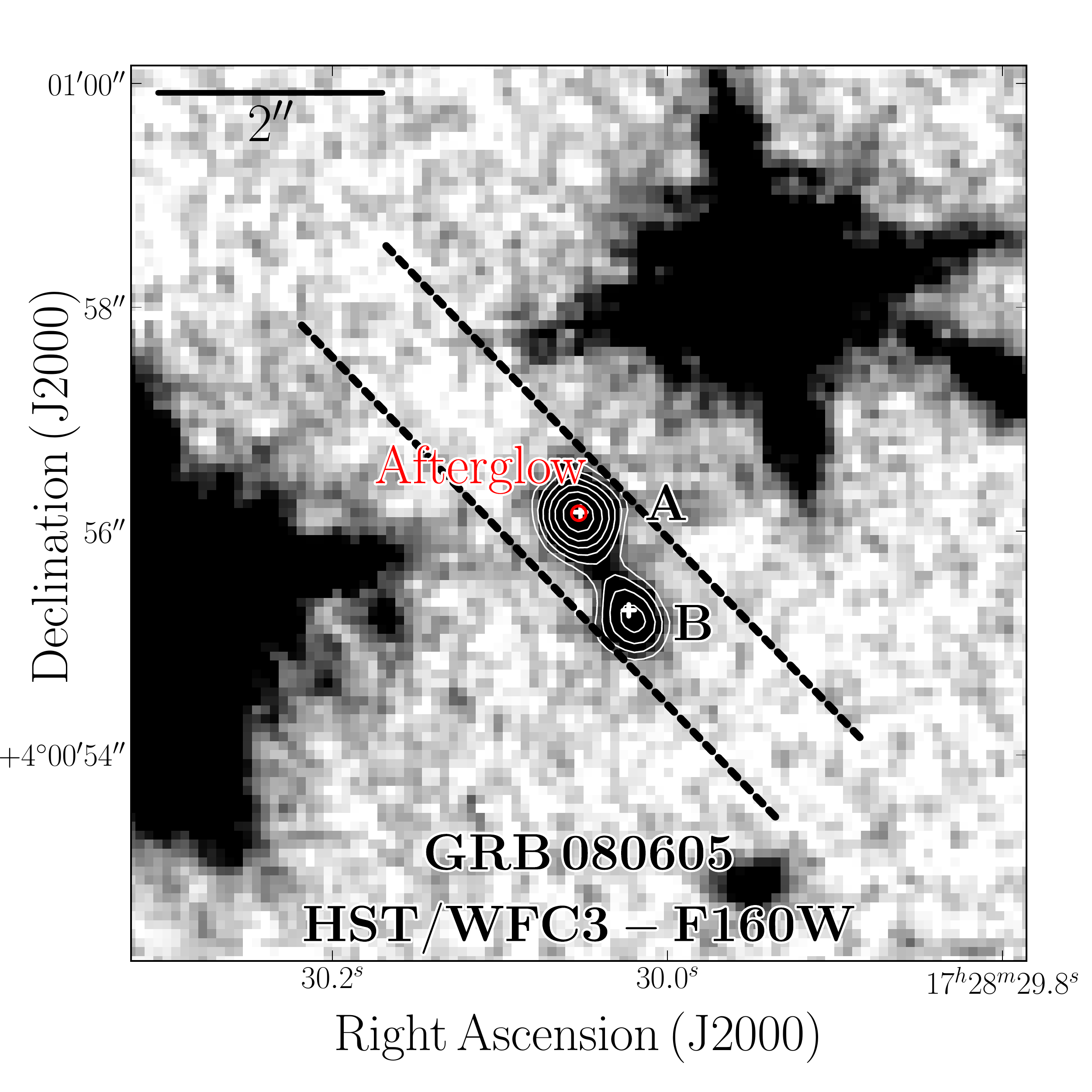}
\caption{{Finding chart ($8\arcsec\,\times\,8\arcsec$) for the host of GRB~080605 as imaged with HST/WFC3. The afterglow position and its uncertainty are indicated by a red circle, and the different components are labeled A and B. The barycenter of each component is indicated by a white cross. The geometry of X-shooter's UVB slit with width of 1\farc{0} is illustrated by dashed black lines. The VIS and NIR slit have the same orientation but a width of 0\farc{9}. Logarithmically spaced contours are shown in white lines.}}
\label{fig:fc}
\end{figure}

{The host of GRB~080605 was observed with the Hubble Space Telescope (HST) and Wide Field Camera 3 (WFC3) as part of a snapshot program targeting GRB hosts (PI: A.~J. Levan, Proposal ID: 12307) on 2012-02-22. HST imaging (see Figure~\ref{fig:fc}) was obtained in the F160W filter in a three-point dither pattern resulting in a total exposure time of 1209~s. Individual images (pixel scale 0\farc{128}/px.) were drizzled to an output image with a pixel scale of 0\farc{08} per pixel. Using several unsaturated stars in the field of view we measure a FWHM of the stellar PSF of $2.6\pm0.1$~px, which is $0\farc{21}\pm0\farc{01}$.}

{To accurately locate the position of the afterglow within its host, we first used a GROND afterglow image from \citet{2011A&A...526A..30G} and calibrated it astrometrically against $\sim80$ sources from the USNO catalog. This sets the absolute astrometric scale with an accuracy of around 0\farc{4} in each coordinate. The uncertainty introduced by centroiding errors of the afterglow is $\approx15$~mas. Afterwards, we registered a deeper GROND host image \citep{2011arXiv1108.0674K} against the afterglow image using common field stars. The mapping uncertainty between the two GROND images is 20~mas. Finally, we used fainter stars from the host image that are unsaturated in the WFC3 frame to tie the space- to the ground-based imaging. In the last step the RMS-scatter of stellar positions is 60~mas in each coordinate, which dominates the total relative accuracy (65~mas) of the position of the afterglow within its host.}

{The host of GRB~080605 is clearly extended in the N/E direction in the HST imaging, and consists of two, somewhat blended components A and B (Fig.~\ref{fig:fc}) with a projected distance of 1\farc{0} between the brightest pixel of each component (corresponding to 8.6~kpc at $z=1.641$). Photometry (see Table \ref{tab:photobs}) was derived using elliptical Kron magnitudes via Sextractor \citep{1996A&AS..117..393B}, an aperture correction of $6\pm4$\% to the total flux \citep{2005PASA...22..118G} and the tabulated HST/WFC3 zeropoints\footnote{\texttt{http://www.stsci.edu/hst/wfc3/phot\_zp\_lbn}} from March 06, 2012. Deblending parameters were set one time to measure the integrated flux of the both components to be comparable to the ground based imaging, and a second time to measure the flux contribution of the individual host components (see also Table~\ref{tab:photobs}). Given the small angular separation, the two components are not resolved in our ground-based imaging.}

\subsection{Ground-based  imaging}

The field of the host of GRB 080605 was also imaged with the LIRIS instrument \citep{2004SPIE.5492.1094M} mounted at the 4.2~m WHT. We obtained a total of 0.55~hr of exposures in the $J$ (average FWHM of the stellar PSF is 1\farcs{4}), and 0.70~hr in the $K_s$-band  (average FWHM of the stellar PSF is 1\farcs{0}) at airmasses between 1.3 and 2.0. The data were reduced and photometry was performed within pyraf/IRAF \citep{1993ASPC...52..173T} in a standard manner. %In the $J$-band a nearby star is blended with the host, which has been psf-subtracted from the original image to perform reliable photometry (see Schulze et al., in preperation for a description of the process). %\citep[see][for a description of the process]{}. 
%Photometry was performed using standard techniques within IRAF. 
Absolute calibration was obtained against roughly $40$ field stars with magnitudes from the 2MASS catalog. This procedure resulted in an absolute photometric accuracy of around 0.05~mag in the $J$, and 0.07~mag in the $K$~band, which is negligible compared to the error introduced by photon statistics. The LIRIS photometry is summarized in Table~\ref{tab:photobs}. 

\begin{table}
\begin{center}
\caption{Photometric measurements. \label{tab:photobs}}
% Trigger: 06:47
%\vspace{-0.3cm}
\begin{tabular}{cccc}
\hline
\noalign{\smallskip}
Instrument & Filter & Exposure (s) &  Brightness (mag)$^{(a)}$ \\  
\hline
HST/WFC3 & F160W & $1209$  &  (A) 22.38 $\pm$ 0.05 \\
" & " & "  &  (B) 23.13 $\pm$ 0.06 \\
" & " & "  &  (A \& B) 21.96 $\pm$ 0.04 \\
LIRIS & $J$ & $1980$  &  (A \& B) 22.2 $\pm$ 0.3 \\
LIRIS & $K_s$ & $2520$  &  (A \& B) 21.8 $\pm$ 0.3\\
\hline
\hline
\end{tabular}
\end{center}
\noindent{

$^{(a)}$ All magnitudes are in the AB system and corrected for a Galactic foreground extinction corresponding to a reddening of $E_{B-V} = 0.137$~mag \citep{1998ApJ...500..525S}.
}

\end{table}

\subsection{X-shooter optical/NIR spectroscopy}
\label{Xsred}
X-shooter \citep{2006SPIE.6269E..98D, 2011arXiv1110.1944V} at the VLT observed the host of GRB 080605 starting at 08:22 UT on 2011-04-26 for a total exposure time of 0.98~hr in the ultra-violet/blue (UVB), 1.01~hr in the visual (VIS), and 1.00~hr in the NIR arm, respectively. Spectroscopy was obtained with slit widths of 1\farcs{0} (UVB), and 0\farcs{9} (VIS and NIR), which results in resolving powers of  $\lambda/\Delta\lambda \approx 5100$, 8800 and 5100 for the three arms. The geometry of the slit is illustrated in Figure~\ref{fig:fc}.

Sky conditions were clear with an average seeing of 1\farcs{2}. In total, four nodded exposures in the sequence ABBA were obtained. In each nodding position a single UVB and VIS frame (885 and 910~s exposure time each), and three NIR frames (300~s exposure time each) were taken. {Data were reduced with the X-shooter pipeline {v. 1.5.0} \citep{2006SPIE.6269E..80G} in physical mode, and the spectra were extracted using an optimal, variance-weighted method in IRAF \citep{1993ASPC...52..173T}. }

The wavelength-solution was obtained against ThAr arc-lamp frames leaving residuals of around 0.2~pixel which corresponds to 6 km s$^{-1}$ at 10\,000~\AA. Flux-calibration was performed against the spectro-photometric standard LTT7987\footnote{\texttt{http://www.eso.org/sci/facilities/paranal/\\instruments/xshooter/tools/specphot\_list.html}} observed during the same night at 09:46 UT, immediately after the science exposures.

The stellar continuum of the host of GRB~080605 is detected in the X-shooter spectrum with a S/N $\approx$ 0.3-0.9 per pixel in parts of the UVB (3600~\AA~to 5500~\AA) and VIS (5600~\AA~to 9800~\AA). {Within the NIR arm the continuum is only marginally seen in the $J$ and $H$ bands with S/N $\sim$ 0.1-0.2 per pixel due to X-shooter's lower sensitivity in this wavelength range. The host is undetected in the wavelength range of the $K$-band with a S/N smaller than $\sim 0.1$ per pixel.}

{A robust flux-calibration within the broad wavelength range of X-shooter's sensitivity is challenging. %Given the spatial extent of the galaxy, the slit widths of 1\farcs{0} for the UVB arm and 0\farcs{9} for the VIS and NIR arms, %, and median seeing of around 1\farcs{2} during the observation, 
%these are expected to be significant. 
We hence further corrected the flux-calibrated X-shooter spectrum in the UVB and VIS arms by integrating it over the filter curves of GROND \citep{2008PASP..120..405G} and HST and matching it to the available host photometry \citep{2011arXiv1108.0674K}. This procedure results in scaling factors of around $1.63\pm0.09$ for the $g'$-band in the UVB arm ($\approx 4590~\AA$), and $1.56\pm0.13$,  $1.35\pm0.12$ and $1.26\pm0.14$ for the $r^{\prime}$, $i^{\prime}$ and $z^{\prime}$ band at $6220$, $7640$~and $8990~\AA$,~respectively. For the NIR arm, we derive factors of $1.4\pm0.4$ for the $J$-band and $1.4\pm0.2$ for the F160W-band. Due to the non-detection of the continuum in the $K$-band, no correction can be obtained between 18\,000 and 23\,000~\AA, but no emission lines are detected in this wavelength regime.}

 {We further tested the absolute flux calibration and its inter- and intra-arm continuity via observations of telluric standard stars taken on the same night. We find that the absolute flux of the telluric is typically recovered within uncertainties of 30\%, while its spectral shape is robust to an accuracy better than 15\% within each arm.}

 %After excluding regions of skylines, the X-shooter data were binned to yield a S/N of approximately 10. The effective width of these bins ranges between roughly 100~\AA~to 200~\AA~and they were used as pseudo-filters in SED fitting of the stellar continuum. Details on the spectral bins and the spectro-photometry are reported in Table~\ref{tab:photobs}.

%assuming a radially symmetric distribution of the host light, and using the spatial distribution of the brightest emission line (H$\alpha$) as a tracer of the galaxy's extend. This profile was propagated in wavelength space from the observed location of H$\alpha$ of 17\,332~\AA~using a standard description of the dependence of the seeing with respect to wavelength ($\rm{FWHM} \propto \lambda^{-1/5}$). This results in wavelength-dependent slit-loss correction factors for example for H$\alpha$ of 1.7 and 1.8 for \oii, respectively. We stress that due to the inherent uncertainties in the radial distribution of host light, and variable seeing conditions during the observations, the estimates on the slit-loss correction factors have errors of at least 20\%.

%Differential atmospheric absorption between the science and standard star observations is small, being $<0.02$~mag throughout the wavelength range of interest (9000-18\,000~\AA), and taken into account during flux calibration.

\section{Results}

\subsection{Host galaxy system and afterglow position}

{The system hosting GRB~080605 consists of two components A and B (see Fig.~\ref{fig:fc}) with barycentric coordinates of RA~(J2000) = 17:28:30.05, decl.~(J2000) = +04:00:56.2 for component A and RA~(J2000) = 17:28:30.02, decl.~(J2000) = +04:00:55.3 for component B, respectively. The half-light radii $r_{\rm{e}}$ in the observed F160W-band (rest-frame $\sim5800\,\AA$) for the two components are marginally resolved ($r^{\rm A}_{\rm{e,5800\,\AA}}\sim0\farc{19}$ or 1.6~kpc, $r^{\rm B}_{\rm{e,5800\,\AA}}\sim0\farc{26}$ or 2.2~kpc). The half-light radius for the total host complex is $r_{\rm{e,5800\,\AA}}=0\farc{41}$ or 3.5~kpc. 
}

{The afterglow position coincides with the center of component A (Figure~\ref{fig:fc}). Within our astrometric accuracy of 65~mas, no significant offset is detected and we conclude that the GRB exploded within a projected distance of 900~pc (90\% confidence) to the central region of component A.}

\subsection{Emission line profile}

\begin{figure}
\centering
\includegraphics[width=0.99\columnwidth]{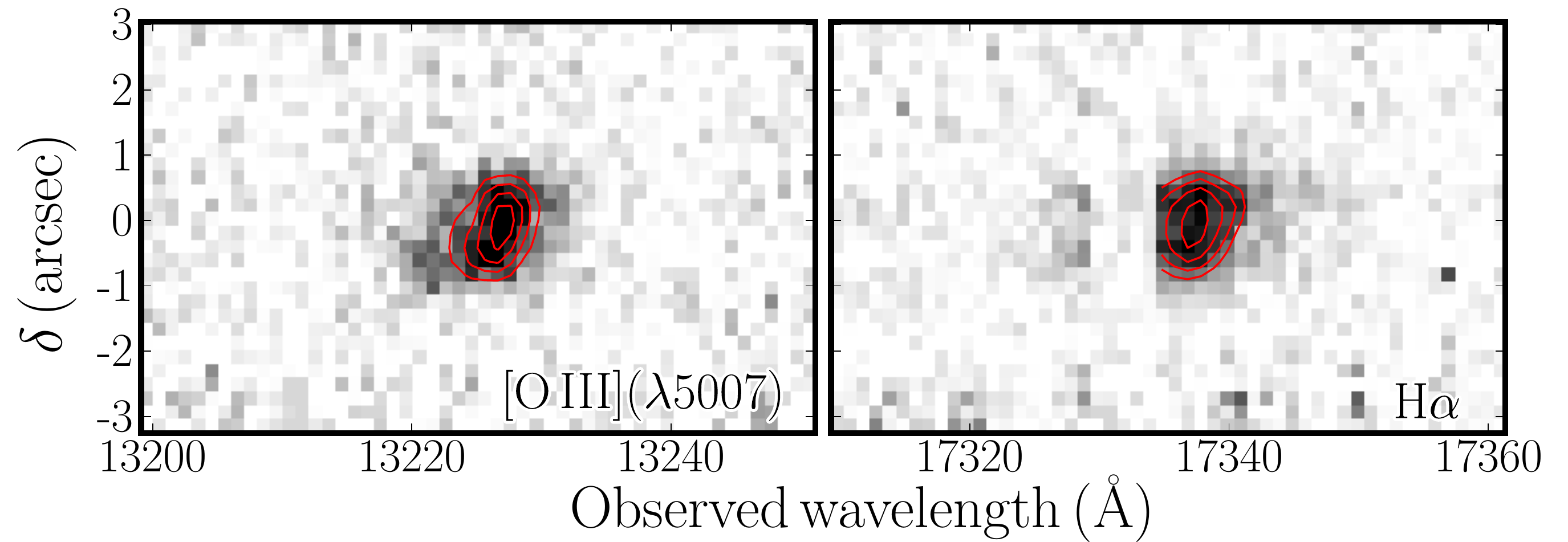}
\caption{{Two-dimensional cutouts of the X-shooter NIR spectrum centered on the observed wavelength of \oiii($\lambda 5007)$ and \ha. Skylines are indicated with grey shading. Linearly spaced contours are shown in red lines.}}
\label{fig:twod}
\end{figure}

The X-shooter spectrum of the host galaxy of GRB 080605 covers the wavelength range between $3050$ and $23\,000~\AA$ (rest-frame $1150$ and $8700~\AA$) and is rich in emission lines. The emission lines are identified as the doublets of \oii, \oiii, \sii, \nii, as well as H$\alpha$, H$\beta$, and \oi.  {The significance of the detection of the Balmer lines, \oiii, \oii~and \nii~($\lambda\,6584$) is $>\, 8\sigma$, while it is between 2 and 4$\sigma$ for \nii~($\lambda\,6548$), the \sii~doublet, and \oi~($\lambda\,6366$).}

{The two emission lines detected at the highest S/N (\oiii~($\lambda5007$) and H$\alpha$, see also Section~\ref{EmissionLines}) are marginally tilted, reflecting the contributions of component A and B. Figure~\ref{fig:twod} shows the two-dimensional cutouts centered at the wavelength of the \oiii~($\lambda5007$) and \ha~lines. They define heliocentric\footnote{The heliocentric correction in the direction of GRB~080605 is 19~km~s$^{-1}$ for our observations.} redshifts of $z_{\rm A}=1.64104\pm0.00004$ and $z_{\rm B}=1.64083\pm0.00007$ measured from the peak of the emission lines. These values correspond to a separation of $\Delta v \sim 20\,\rm{km\,s^{-1}}$ (Figures~\ref{fig:fc} and \ref{fig:twod}).}

{For the fainter emission lines, we lack signal-to-noise ratio in our X-shooter spectrum and individual contributions of components A and B are strongly blended and can not be resolved. Similar to the ground-based photometry, we will thus report line-fluxes integrated over the complete host galaxy complex (Section~\ref{EmissionLines}) in the following.}

\subsection{Host SED}

\label{SED}
{Fitting the HST and LIRIS NIR photometry of the entire host system together with published broad-band magnitudes \citep[see][for details]{2011arXiv1108.0674K} in LePhare\footnote{\texttt{http://www.cfht.hawaii.edu/$\sim$arnouts/lephare.html}} \citep{1999MNRAS.310..540A, 2006A&A...457..841I} yields the galaxy parameters listed in Table~\ref{tab:hostprop}. Here we assumed models from \citet{2003MNRAS.344.1000B} based on an initial mass function (IMF) from \citet{2003PASP..115..763C} and a Calzetti dust attenuation law \citep{2001PASP..113.1449C}. Given that both method and data are largely unchanged, these values are only slightly refined with respect to those computed by \citet{2011arXiv1108.0674K}.}
\begin{table}
\begin{center}
\caption{Host parameters from stellar population synthesis modeling. \label{tab:hostprop}}
% Trigger: 06:47
%\vspace{-0.3cm}
\begin{tabular}{cc}
\hline
\noalign{\smallskip}
Absolute magnitude $M_B$ (mag$_{\rm AB}$) & $-22.4\pm0.1$ \\ 
Age (Gyr) &  $0.19_{-0.10}^{+0.09}$ \\  
Effective reddening $E_{B-V}^{\rm{stars}}$ (mag) & $0.10_{-0.08}^{+0.05}$  \\ 
$M_\ast$ ($10^{9}M_{\sun}$) & $8.0^{+1.3}_{-1.6}$ \\ 
SFR$_{\rm{SED}}$ ($M_{\sun}\,\rm{yr}^{-1}$) & $49_{-13}^{+26}$ \\ 
sSFR$_{\rm{SED}}$ (Gyr$^{-1}$) &  $6_{-2}^{+5}$ \\ 
\hline
\hline
\end{tabular}
\end{center}
\end{table}

%\subsection{Host galaxy spectrum}

\subsection{Emission line fluxes}
\label{EmissionLines}

\begin{figure*}
\centering
\includegraphics[width=1.5\columnwidth]{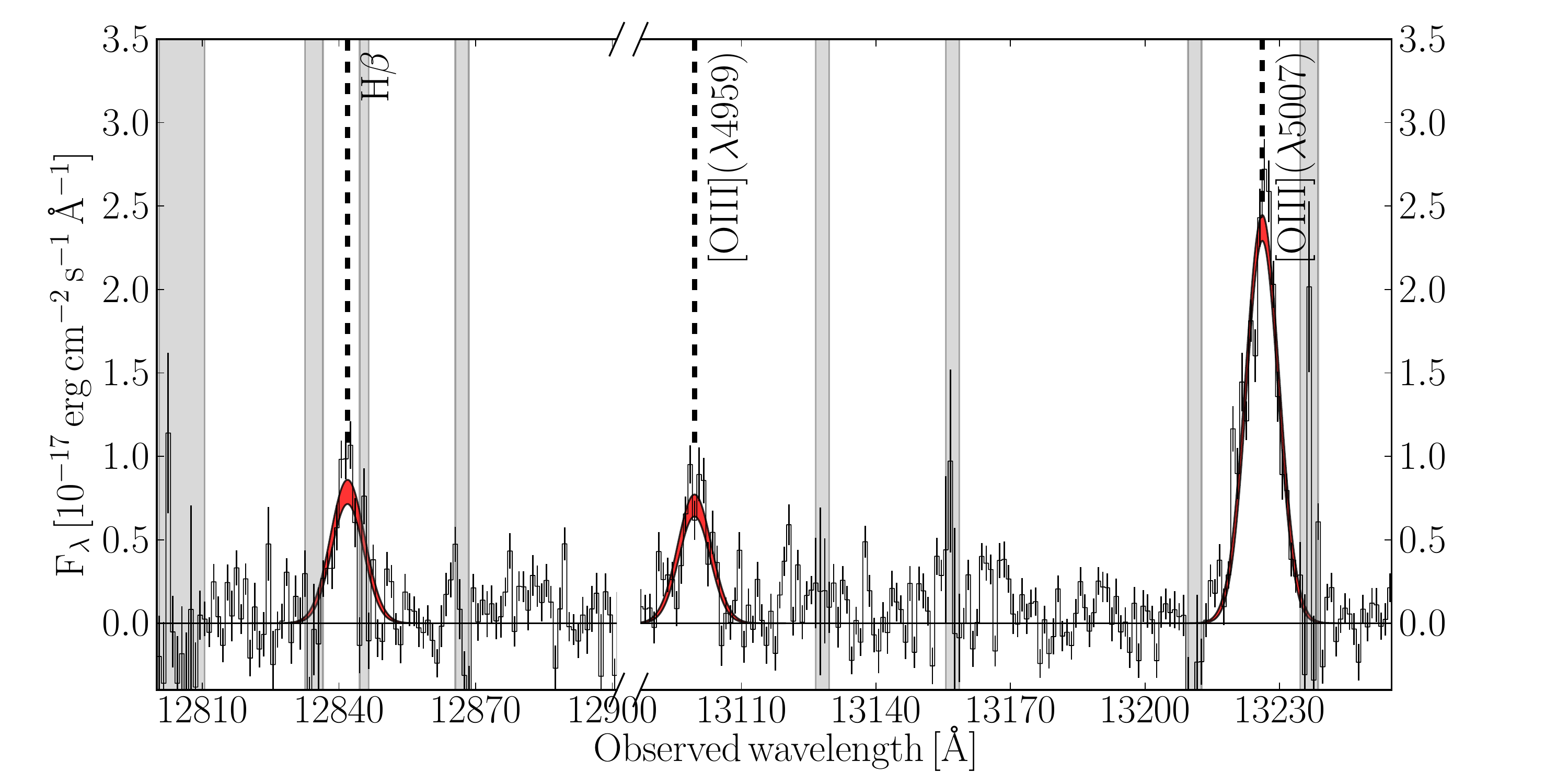}\\
\includegraphics[width=1.5\columnwidth]{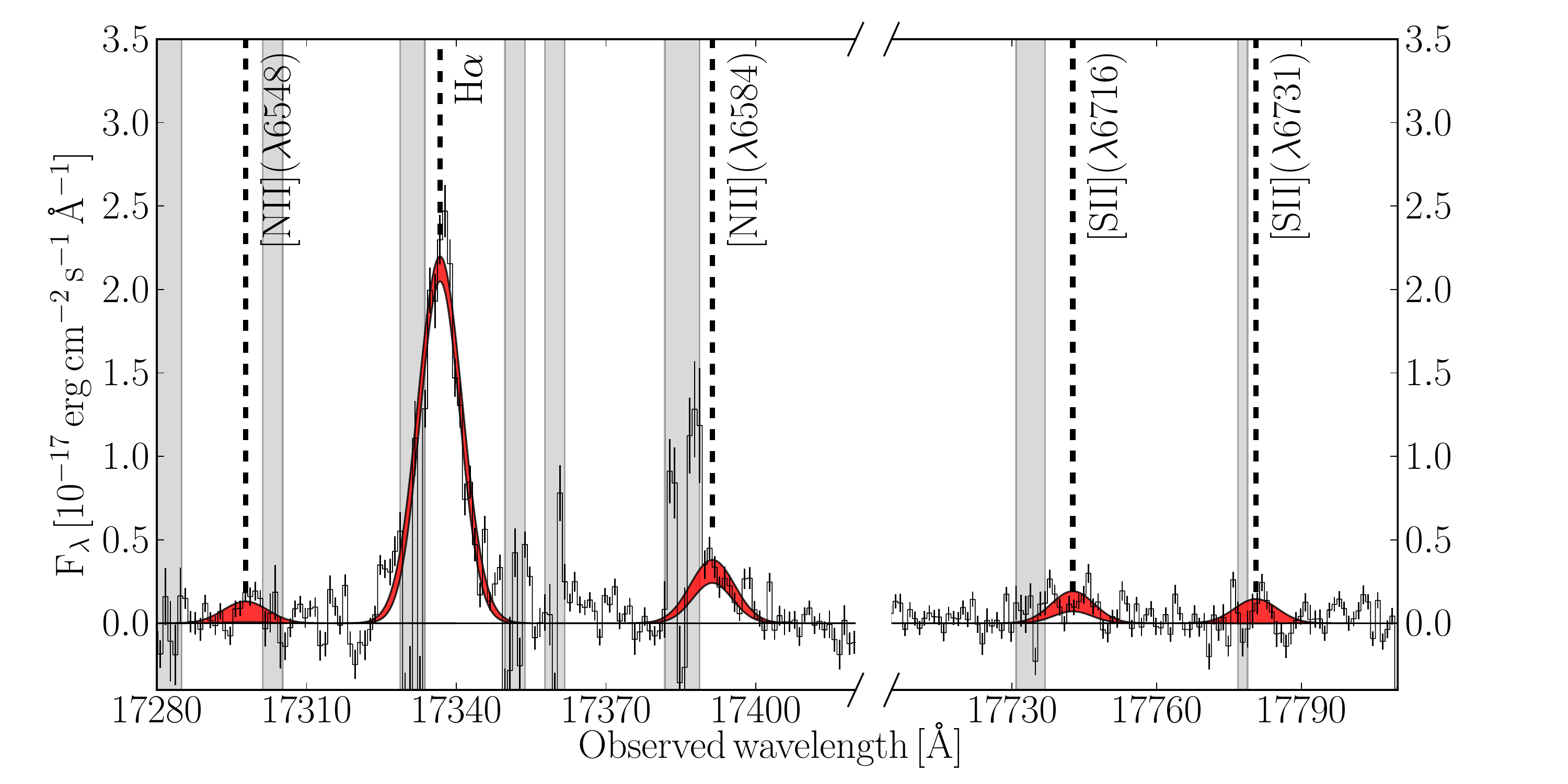}
\caption{Continuum-subtracted emission lines used to determine the gas-phase metallicity in the X-shooter spectrum, as well as the \sii~doublet. The black line shows the raw spectrum including errors, and the red-shaded areas denote the 90\% confidence region of the fit of the emission lines using Gaussians. Grey shaded areas denote wavelength regions that have been omitted in the fitting due to skyline contamination.}
\label{fig:eml}
\end{figure*}

{In the measurement of emission line fluxes (Table~\ref{tab:eml}), the redshift (i.e., line centroids) and line widths were fitted simultaneously by tying weak emission lines to those detected at high S/N. In detail, we linked the parameters of the two components of the \oii~doublet in the visual arm, as well as the Gaussian widths and centroids of the various emission lines in the NIR arm. Although the emission of the forbidden and recombination lines does not necessarily arise from the same physical components, the assumption of a common redshift and line width provides a fair approximation and a good fit to the data (Fig.~\ref{fig:eml}). The robustness of the procedure is further supported by, within errors, unchanged line parameters, fluxes and flux ratios when using different combination of ties (i.e., free FWHM, tying H$\beta$ to \oiii$\,(\lambda$5007) or all lines except \oii~to each other) {or allowing for multiple Gaussians components in the individual lines}.}

{In addition, we cross-checked our method by numerically integrating the flux of the emission lines. Here, errors were estimated via Monte-Carlo techniques. This results in values that are consistent with those of the Gaussian fitting at $2 \sigma$ confidence, but is more sensitive to skylines and small-scale irregularities in the data. It further disregards the physical information of a common redshift, and hence results in larger errors than the Gaussian fitting in particular for lines with low S/N, or those affected by skylines. % Emission line fitting based on (a combination of) Gaussian lines is common also for X-shooter data \citep{2010MNRAS.402.2335P, 2011A&A...533A..15D}, 
We thus report fit-based values in Table~\ref{tab:eml}. Our conclusions remain unchanged when using different Gaussian fitting methods or numerical integration techniques for the line flux measurements.}

From the observed FWHM of the \oii~doublet ($\approx$ 5~\AA), and assuming a resolving power of 8800 of X-shooter's VIS arm, we derive a measured velocity dispersion $\sigma$ of around $\sigma \sim $ 50~km~s$^{-1}$ for the host galaxy complex, comparable to star-forming systems of similar mass observed through gravitational lenses \citep[e.g.,][]{2010MNRAS.406.2616C}.

We do not detect significant emission from the resonant Ly$\alpha$ transition. Using the redshift, and assuming an intrinsic FWHM of twice the recombination lines \citep[e.g.,][]{2010MNRAS.408.2128F}, we set a limit on the Ly$\alpha$ flux of $4.7\times10^{-17}\,\rm{erg}\,\rm{cm}^{-2}\,\rm{s}^{-1}$ ($7.6\times10^{-17}\,\rm{erg}\,\rm{cm}^{-2}\,\rm{s}^{-1}$ after matching the spectrum to photometry) at the redshifted Ly$\alpha$ wavelength of $3210~\AA$. It is estimated from an artificial emission line added on top of the sky contribution at the respective wavelength range, folded with the error spectrum and represents the flux that is detected at a combined S/N~$>$~3 in 99\% of all iterations. Similar limits are obtained when allowing for an offset of several ten to few hundreds of $\rm{km}\,s^{-1}$ for Ly$\alpha$ with respect to the recombination lines \citep{Bo2012}. The non-detection of Ly$\alpha$ is further discussed in Section~\ref{LyA}. 

In the further analysis, we matched the spectrum to broad-band photometry (see Section \ref{Xsred}), and applied when appropriate the correction for an average stellar Balmer absorption using a rest-frame equivalent width of {1~\AA~\citep{2008ApJ...686...72C, 2011ApJ...730..137ZF}}, and for host galaxy extinction using the Balmer decrement (see Section \ref{Balmer}). The corresponding wavelength-dependent factors are shown in Table~\ref{tab:eml}.

\begin{table}
\centering
\caption{Emission lines in the X-shooter spectrum. \label{tab:eml}}
% Trigger: 06:47
%\vspace{-0.3cm}
\begin{tabular}{cccc}
\hline
\noalign{\smallskip}
Transition & Wavelength$^{(a)}$ & Flux$^{(b)}$ & Correction$^{(c)}$ \\  
\hline
Ly$\alpha$ & 1215 & $< 4.7$  & $3.2_{-1.6}^{+7.8}$ \\
\oii &  3727  & $10.6\pm0.6$ & $1.7_{-0.5}^{+1.3}$  \\
\oii & 3730 & $12.2\pm0.9$ & $1.7_{-0.5}^{+1.3}$ \\
H$\beta$ & 4863 & $7.2\pm0.5$ & $1.8_{-0.6}^{+1.1}$ \\
\oiii & 4960 & $6.5\pm0.4$ & $1.8_{-0.6}^{+1.0}$  \\
\oiii & 5008 & $21.6\pm0.6$  & $1.8_{-0.6}^{+1.0}$ \\
\oi & 6366 & $1.6\pm0.4$  & $1.7_{-0.4}^{+0.7}$ \\
\nii & 6550 & $0.8\pm0.4$ & $1.6_{-0.3}^{+0.6}$ \\
H$\alpha$ & 6565 & $22.4\pm1.0$ & $1.6_{-0.3}^{+0.6}$ \\
\nii & 6585 & $3.2\pm0.4$ & $1.6_{-0.3}^{+0.6}$ \\
\sii & 6718 & $1.2\pm0.4$ & $1.6_{-0.3}^{+0.6}$ \\
\sii & 6733 & $0.9\pm0.4$ & $1.6_{-0.3}^{+0.6}$ \\
\hline
\hline
\end{tabular}
\noindent{\begin{flushleft}$^{(a)}$ Rest-frame vacuum wavelength in units of $\AA$.

$^{(b)}$ Galactic extinction corrected flux in units of $10^{-17}$~erg~s$^{-1}$~cm$^{-2}$. The flux is quoted as measured in the X-shooter spectrum (without correction). The flux error is statistical only, and does not contain the error of the absolute flux calibration.

$^{(c)}$ The given correction includes the matching factor to broad-band photometry, stellar Balmer absorption if applicable, and reddening according to the Balmer line ratio. These factors are not independent. In particular it was assumed that \nii~and H$\alpha$, for example, have identical values (except for the Balmer absorption).\end{flushleft}
 }
\end{table}

Comparing the emission line ratios of \oiii/H$\beta$ versus \nii~($\lambda$6584)/H$\alpha$ against standard diagnostic relations \citep[e.g.,][]{2001ApJ...556..121K, 2003MNRAS.346.1055K}, a significant contribution of an AGN to the host emission of GRB~080605 is readily excluded. Measurements of emission line fluxes and upper limits are reported in Table~\ref{tab:eml}.

\subsection{Balmer decrement}
\label{Balmer}

The ratio between the Balmer lines H$\alpha$ and H$\beta$ is a tracer of the visual extinction towards the \ion{H}{ii} regions. We used the respective photometry-matched and stellar Balmer absorption corrected line fluxes to derive the intrinsic Balmer ratio, which is a direct measure of the selective reddening, or the total visual extinction under the assumption of a specific extinction law (and treating the \ion{H}{ii} regions as point-like). This probes a different physical quantity than the reddening value inferred from fitting the galaxy's SED from Table~\ref{tab:hostprop}, as the SED modeling is sensitive to the attenuation of the stellar component which depends on the topology of the ISM and dust and galaxy geometry \citep[e.g.,][]{2004ApJ...617.1022P}.%, and is typically around a factor of two lower than the gas extinction derived from the Balmer decrement \citep[e.g.,][]{2000ApJ...533..682C, 2005MNRAS.358..363C, 2009ApJ...691..182S}.

Under standard assumptions of electron density ($10^2\, \rm{cm}^{-3} \lesssim n_e \lesssim 10^4$~cm$^{-3}$, see also Sect.~\ref{Te}) and temperature ($T_e \sim 10^4$~K) for case B recombination \citep{1989agna.book.....O}, the Balmer ratio indicates an average reddening  towards the \ion{H}{ii} regions of $E_{B-V}^{\rm{gas}} = 0.07_{-0.07}^{+0.13}$~mag. This corresponds to $A_{V}^{\rm{gas}} = 0.22_{-0.22}^{+0.40}$~mag when assuming a MW-type extinction law with $R_V = 3.1$ \citep{1989ApJ...345..245C}. The reddening corrections according to the Balmer decrement for all emission lines except Ly$\alpha$ are fairly robust and independent on the assumption of a specific extinction law, as there is little difference within the wavelength range of the Balmer lines between sight-lines through local galaxies \citep[e.g.,][]{1992ApJ...395..130P} or with respect to extra-galactic extinction laws derived from GRB afterglows \citep{2011arXiv1110.3218S}. 

\subsection{Electron density}
\label{Te}
\begin{figure}
\centering
\includegraphics[width=.8\columnwidth]{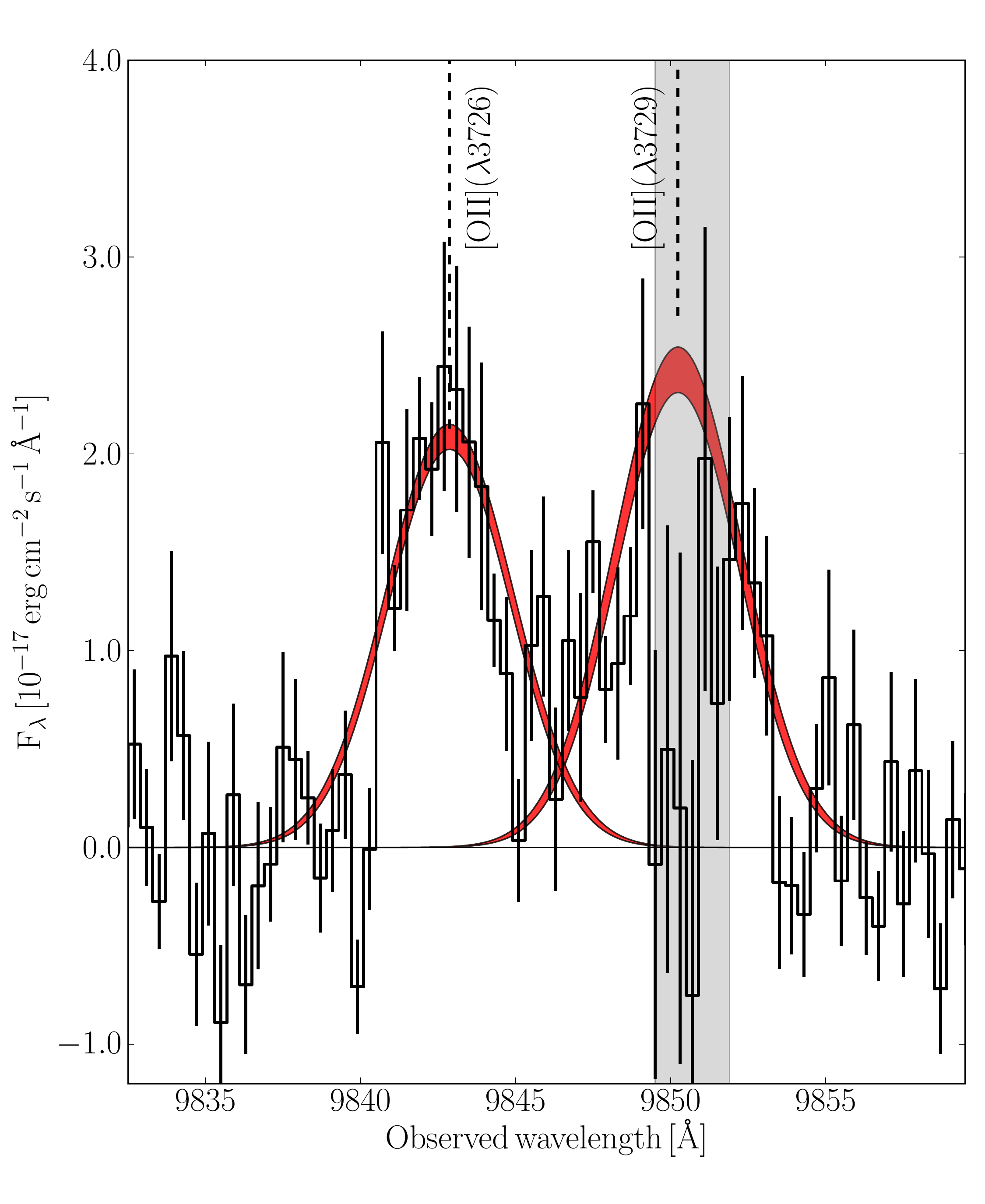}
\caption{The continuum-subtracted \oii~$(\lambda\lambda\, 3726\, 3729)$ doublet. Lines and shadings are the same as in Fig.~\ref{fig:eml}.}
\label{fig:o2eml}
\end{figure}

The flux ratio between the two components of the \oii~doublet is sensitive to the electron density \citep{2006agna.book.....O}. Individual components are resolved and well detected because of the high spectral resolution of X-shooter (see Fig.~\ref{fig:o2eml}). Assuming an electron temperate $T_e$ of $10^4$~K, we derive an electron density of $n_e \sim 200\,\rm{cm}^{-3}$. This value is typical for Galactic \ion{H}{ii} regions \citep[e.g.,][]{2000A&A...357..621C}. The low significance of the detection of the \sii~doublet (see Fig.~\ref{fig:eml}) prevents meaningful constraints on the electron density based on \sii.

\subsection{Star-formation rate}

Emission line fluxes of H$\alpha$ and \oii, as well as the UV continuum flux trace the un-obscured star-formation within a galaxy \citep{1998ARA&A..36..189K, 2004AJ....127.2002K}. The H$\alpha$-derived value presents the most reliable optical indicator of a galaxy's SFR, as it is independent of metal abundances, and less sensitive to the uncertainties in the visual extinction than the other methods. SFR values depend quite strongly on the assumption of the IMF. In the following, we report values based on the initial formulation of the SFR from \citet{1998ARA&A..36..189K}, but converted to a Chabrier IMF \citep{2003PASP..115..763C}\footnote{Assuming a Salpeter IMF would increase all SFR estimates by a factor of $\approx 1.7$ \citep[e.g.,][]{2009ApJ...706.1364F}.}. Based on H$\alpha$, the SFR of the host of GRB~080605 is SFR$_{\rm{H}\alpha} = 31^{+12}_{-6}\,M_{\sun}\,\rm{yr}^{-1}$. Here we used correction-factor and its error (see Tab.~\ref{tab:eml}), and thus include the uncertainty in the flux calibration and host-intrinsic reddening. {SFRs from \oii~(SFR$_{\oii} = 55^{+55}_{-22}\,M_{\sun}\,\rm{yr}^{-1}$) and the SED modeling (SFR$_{\rm{SED}} = 49_{-13}^{+26}\,M_{\sun}\,\rm{yr}^{-1}$) are within the larger uncertainties in good agreement with the H$\alpha$-derived value.}

Optically derived SFRs do not provide a full picture of the total (obscured and unobscured) star-formation in a galaxy. Based on sub-mm and radio measurements, there is evidence that the total SFR of few GRB-selected galaxies can be around or even higher than $100 \,M_{\sun}\,\rm{yr}^{-1}$ \citep[e.g.,][]{2003ApJ...588...99B, 2004MNRAS.352.1073T, Michal2012}. The sample of sub-mm detected GRB hosts, however, is still very limited \citep[e.g.,][]{2008ApJ...672..817M}, and a full census of the actual SFR of GRB hosts will have to await the advent of statistically representative samples observed with sensitive far-infrared, sub-mm or radio observatories such as \textit{Herschel}, ALMA or the eVLA. 

Despite these limitations, optically-derived SFRs  are well-established tools for the characterization of galaxies. We will thus put the host of GRB~080605 into the context of SFRs from field galaxies derived in a similar manner, with the caveat that the reported SFRs might trace only a fraction of the total SFR of a given galaxy.

Together with the stellar mass measurement of $M_* = 8.0^{+1.3}_{-1.6}\times 10^9M_{\sun}$, the specific SFR (sSFR$_{\rm{H}\alpha}$ = SFR$_{\rm{H}\alpha}$/$M_{*}$) and growth timescale $\tau=1/\rm{sSFR_{\rm{H}\alpha}}$ are 4~Gyr$^{-1}$ and 260~Myr, respectively, making the host of GRB~080605 a highly active and star-bursting galaxy.

\subsection{Metallicity}

The gas-phase metallicity of galaxies is typically measured using different diagnostic ratios of emission lines originating in \ion{H}{ii} regions \citep[see e.g.,][and references therein]{2008ApJ...681.1183K}. Most commonly used are the $R_{23}$ calibrator, that requires measurements of line fluxes from \oii, \oiii~and H$\beta$ \citep{1979MNRAS.189...95P, 1991ApJ...380..140M, 2004ApJ...617..240K}, the O3N2 and N2 diagnostics which uses ratios of \oiii, and H$\beta$,  and/or \nii~and H$\alpha$ \citep{1979A&A....78..200A, 2004MNRAS.348L..59P}, and the N2O2 indicator via \oii~and \nii~\citep{2002ApJS..142...35K}. For a detailed description on the individual strong-line diagnostics we refer to \citet{2008ApJ...681.1183K}.

The $R_{23}$ method is double-valued but its degeneracy can be broken via the line ratios of {\nii/H$\alpha$ or $\nii/\oii$. In our case, \nii/H$\alpha=0.14\pm0.02 $ and $\nii/\oii =0.10\pm0.04$.}
The significant flux detected in \nii~strongly points to the upper branch solution. Similarly, the N2O2 ratio is only applicable for high metallicities with $\log(\nii/\oii) > -1.2$. Due to the large difference in wavelength of the lines used by N2O2 and $R_{23}$, their values are sensitive to the reddening in the host and wavelength-dependent errors in the flux calibration. 

Both the O3N2 (see Eq.~1) and N2 methods, however, use flux ratios of adjacent emission lines, which are relatively close in wavelength space, and in our case are all within the NIR arm. The observed lines of \oiii~and H$\beta$ are located in the $J$, and \nii~and H$\alpha$ in the $H$-band. Errors in the flux calibration or systematic uncertainties due to flat-fielding, slit-losses, intrinsic host extinction or reddening in the Galaxy are hence not going to affect the overall metallicity measurement in this case. 

%In the calculation of the metallicity for the different diagnostic ratios, we treat possible errors in the line fluxes ratios as well the uncertainty in the reddening rigorously. Specifically, we used the the appropriate correction factors from Table~\ref{fig:eml} in the analysis and propagated errors accordingly. %The combined correction factor was then propagated in wavelength space and applied in the estimation of the metallicity error.

Based on O3N2, for example, the oxygen abundance is \citep{2004MNRAS.348L..59P}:

\begin{equation}
12 + \log(\rm{O}/\rm{H}) = 8.73-0.32\times\log\left({\frac{F_{\oiii(\lambda5007)}/F_{\rm{H}\beta}}{F_{\nii(\lambda6584)}/F_{\rm{H}\alpha}}}\right)
\end{equation}
{which is $12 + \log(\rm{O}/\rm{H}) = 8.31\pm0.02$ for GRB~080605. Using the N2, $R_{23}$ and N2O2 strong-line indicators, the oxygen abundance is $12 + \log(\rm{O}/\rm{H}) = 8.36\pm0.03$ for N2, $12 + \log(\rm{O}/\rm{H}) = 8.45^{+0.09}_{-0.12}$ for $R_{23}$, and $12 + \log(\rm{O}/\rm{H}) = 8.60^{+0.11}_{-0.19}$ for N2O2. Here, all errors are based on the uncertainties of the line flux measurement and the correction factor only (see Table~\ref{tab:eml}), and do not include the systematic error inherent to the calibrator. Errors correctly reflect the larger uncertainty in the strong-line diagnostics that include flux ratios between lines in different wavelength regimes ($R_{23}$ and N2O2). Oxygen abundances based on different indicators are further summarized in Table~\ref{tab:met}. We use the appropriate diagnostics when comparing to literature values.}
 
\begin{table*}
\centering
\caption{Oxygen abundances based on different strong-line indicators \label{tab:met}}
% Trigger: 06:47
%\vspace{-0.3cm}
\begin{tabular}{cccccc}
\hline
\noalign{\smallskip}
Indicator & Lines / Methods & $12 + \log(\rm{O}/\rm{H}) $ &  $Z/Z_{\sun}$ & Uncertainty$^{(a)}$ (dex) & References$^{(b)}$ \\  
\hline
$R_{23}$ & \oii (3727), \hb, \oiii (5007) & $8.45^{+0.09}_{-0.12}$ & $0.63 \pm 0.15$ & 0.15 & (1), (2), (3) \\  
  & \oii (3727), \hb, \oiii (5007) &  $8.50^{+0.10}_{-0.13}$ & $0.64\pm0.17$  & $\sim$~0.1 & (4, 6) \\  
O3N2 & \hb, \oiii (5007), \ha, \nii (6584) &  $8.31\pm0.02$ & $0.42\pm0.02$  & 0.14  & (5) \\ 
         & \oiii (5007), \nii (6584) & $8.46\pm0.10$ & $0.59^{+0.15}_{-0.12}$  & 0.24 & (6) \\ 
N2 & \ha, \nii (6584) &  $8.36\pm0.03$ & $0.47\pm0.04$  & 0.18 & (5) \\ 
    & \ha, \nii (6584) &  $8.52\pm0.06$ & $0.68^{+0.09}_{-0.08}$  & 0.12 & (6) \\ 
N2O2 & \oii (3727), \nii (6584) &  $8.60^{+0.11}_{-0.19}$ & $0.82^{+0.24}_{-0.29}$ & $\sim$~0.1 & (7) \\ 
 & \oii (3727) \nii (6584) &  $8.53^{+0.14}_{-0.24}$ & $0.69^{+0.27}_{-0.30}$  & 0.10 & (6) \\  
 Combined fit & $R_{23}$, O3N2, N2, N2O2 &  $8.52\pm0.09$ & $0.68^{+0.15}_{-0.13}$ & included & (4), (6), (8)  \\
\hline
\hline
\end{tabular}
\noindent{\begin{flushleft}$^{(a)}$ Systematic 1$\sigma$ scatter inherent to the diagnostic line ratio.

$^{(b)}$ References for the indicator: (1) \citet{1979MNRAS.189...95P}; (2) \citet{1991ApJ...380..140M}; (3) \citet{2004ApJ...617..240K}; (4) \citet{2008A&A...488..463M}; (5) \citet{2004MNRAS.348L..59P}; (6) \citet{2006A&A...459...85N}; (7) \citet{2002ApJS..142...35K}; (8) \citet{2011MNRAS.414.1263M}. \end{flushleft} 
 }
\end{table*}

\section{Discussion}

The metallicity of GRB hosts is measured either directly in absorption using the bright afterglow emission, or, as in this work, in emission via host galaxy spectroscopy and strong-line diagnostics. In the former case, measurements are typically restricted to $z \gtrsim 2$ \citep[e.g.,][]{2006A&A...460L..13J, 2006NJPh....8..195S}, while the latter case requires NIR spectroscopy for $z \gtrsim 1$. Emission line metallicities are calibrated on local samples, and hence depend on the assumption that the physical processes underlying these diagnostic ratios are still valid at high redshift. {There is hence considerable systematic uncertainty between metallicities derived directly in absorption or through emission lines. Furthermore, metallicity measurements at high-redshift via the different techniques are available for only a few objects. They tend to agree reasonably well \citep[see e.g.,][]{2010A&A...510A..26D}, but systematic effects in a direct comparison remain hard to quantify until larger samples of objects with both, ISM as well as gas-phase metallicities, become available.}

Our measurement of the gas-phase metallicity of the host of GRB~080605 represents a first view into the metal abundances of GRB hosts in the redshift range $1 \lesssim z \lesssim 2$ (Fig.~\ref{fig:met}). Generally, the distribution in metallicity as inferred from GRB-DLAs shows a large dispersion with values $0.01 \lesssim Z/Z_{\sun}$ \citep{2004A&A...419..927V, 2008ApJ...685..344P, 2010ApJ...720..862R} to solar or even super-solar \citep{2009ApJ...691L..27P, 2003ApJ...585..638S, 2011arXiv1110.4642S}. A similar spread in metallicities is also found in hydrodynamical solutions of individual sight-lines through GRB hosts \citep{2010MNRAS.402.1523P}. The metallicity derived from afterglow spectroscopy could be dominated by sight-line effects, and large samples might be required to assess the general properties of GRB hosts via afterglow spectroscopy in a statistical approach. 

Host-integrated metallicities via emission lines should therefore give a more self-contained picture of the metal-enrichment of the ISM in high-redshift GRB hosts. Galaxy metallicity measurements are however challenging observationally in a stellar mass range around or below $10^{10}\,M_{\sun}$, and thus still sparse, in particular at $z > 1$. Current GRB host samples are furthermore subject to complex selection biases \citep{2011arXiv1108.0674K}, which are only resolved through statistical samples of high completeness \citep[e.g.,][]{2009ApJS..185..526F, 2009ApJ...693.1484C, 2011A&A...526A..30G, 2011arXiv1112.1700S, Jens2012}. Consequently, a consistent picture of the relation between galaxies selected through GRBs and normal field galaxies is not yet reached.

\subsection{The host of GRB~080605 within the sample of GRB hosts}

\begin{figure}
\centering
\includegraphics[width=\columnwidth]{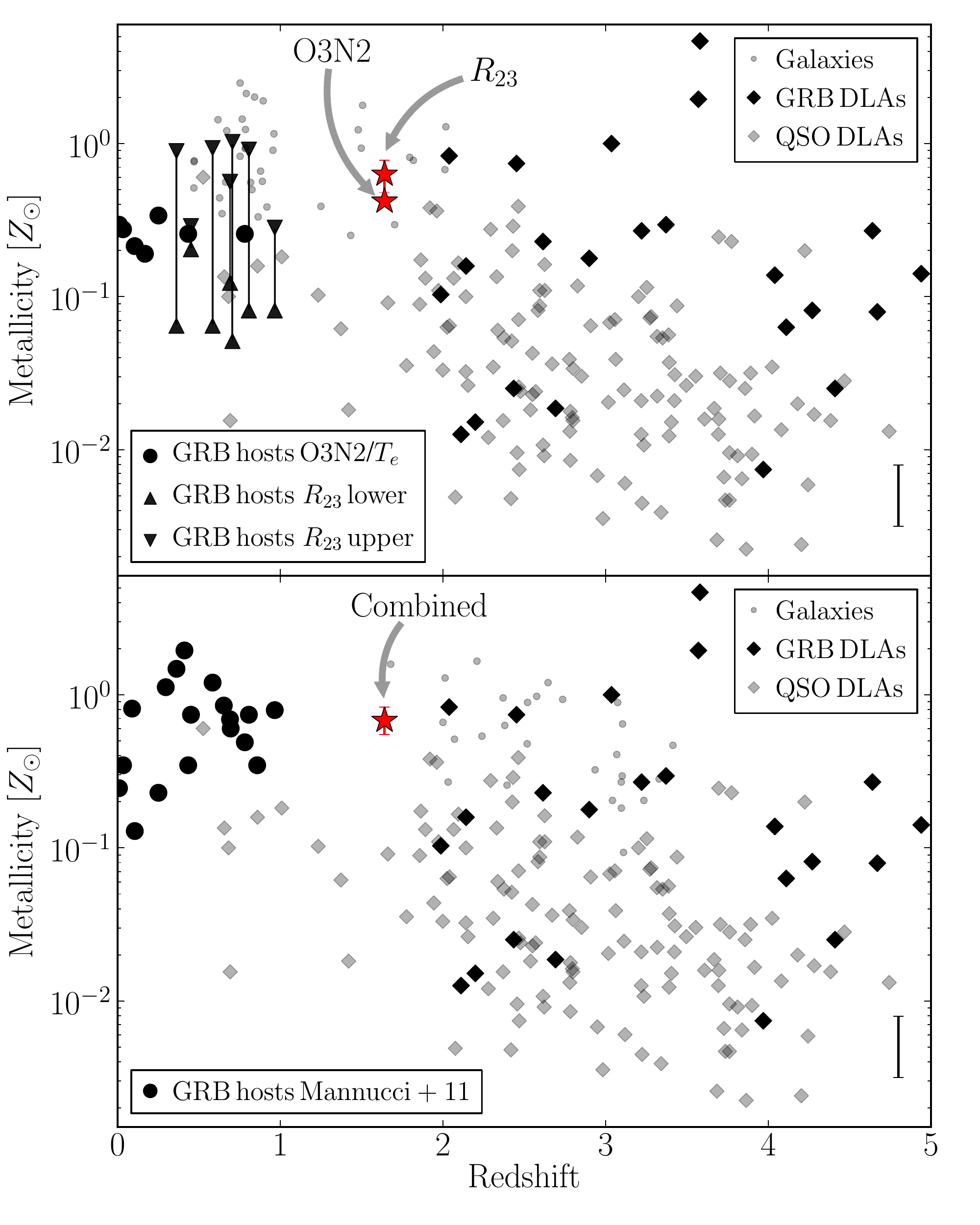}
\caption{{Metallicity of the host of GRB~080605 (star). Other GRB host metallicities are shown with black circles and upward/downward triangles as compiled and in the scale of \citet{2009ApJ...691..182S} (top panel) and \citet{2011MNRAS.414.1263M} (bottom panel).} Field galaxies are shown as grey dots \citep{2005ApJ...635..260S, 2009A&A...495...73P, 2009MNRAS.398.1915M, 2009ApJ...691..140H, 2011MNRAS.413..643R}, and in similar metallicity scales as the GRB measurements. Absorption metallicities from GRB afterglows \citep[][and references therein]{2010ApJ...720..862R,  2010A&A...523A..36D, 2011arXiv1110.4642S, 2011100219A} and QSOs \citep{2003ApJS..147..227P} are plotted as black and grey diamonds, respectively. {Errorbars for individual events in the comparison samples are omitted to enhance clarity.} The error bars at the bottom right corner of each panel illustrate {uncertainties of 0.2~dex., which are typical for both, GRB-DLA \citep[e.g.,][]{2010ApJ...720..862R, 2011100219A} and GRB host metallicity \citep[e.g.,][]{2011MNRAS.414.1263M} measurements}.}
\label{fig:met}
\end{figure}

{With respect to previous GRB host galaxies, the metallicity, stellar mass and star-formation rate of the host of GRB~080605 are relatively high (see Fig.~\ref{fig:met}). With a metallicity around half solar, a stellar mass of $8\times 10^9~M_{\rm{\sun}}$ and SFR $\sim 30~M_{\sun}\,\rm{yr}^{-1}$, it is significantly enriched with metals and vigorously forming stars.} This contradicts the suggestion, that an upper metallicity limit\footnote{Adopted to our reference solar oxygen abundance.}  for cosmological, $z \gtrsim 1$, GRBs of $Z \lesssim 0.2\, Z_{\sun}$ exists \citep{2006AcA....56..333S}. 

The substantial gas-phase metallicity of the host is even more intriguing, as GRB~080605 itself is energetic. The inferred isotropic-equivalent energy release in $\gamma$-rays is $E_{\gamma,\rm iso} \sim 2.2\times10^{53}$~erg as calculated from the prompt emission data from \citet{2008GCN..7854....1G}. This value puts GRB~080605 within the most-energetic 15\% of all \textit{Swift} bursts \citep{2007ApJ...671..656B, 2010ApJ...711..495B}. 

A connection between host metallicity and $\gamma$-ray energy release of the GRB, or a metallicity cut-off might be expected in the collapsar scenario \citep{1993ApJ...405..273W, 1999ApJ...524..262M}, for example. Progenitor stars with lower metallicities are likely to have higher angular momentum due to smaller wind losses, and thus result in a more energetic explosion \citep[e.g.,][]{2005A&A...443..581H, 2005A&A...443..643Y}. An energetic burst such as GRB~080605 would hence be more likely in a low-metallicity environment, in contrast to our observations. Our observations are, however, in line with the work of \citet{2007MNRAS.375.1049W} and \citet{2010ApJ...725.1337L}, who find no correlation between $E_{\gamma, \rm iso}$ and host metallicity in 18 GRBs at $z < 1$. 

{The role of metallicity in long GRB progenitors is thus far from being understood. The metallicity distribution of a representative GRB host sample will indirectly also allow us to put constraints on the metal content of the progenitor. For example, complex scenarios of stellar evolution, or binary models for the formation of long GRBs  \citep[e.g.,][]{1999ApJ...526..152F} can both relax the constraints on progenitor metallicity. In addition, even within a metal-rich galaxy a metal-poor progenitor could in principle form in specific regions of fairly primordial chemical composition such as gas inflows or in (merging) galaxies with substantial diversity in their metal enrichment.} To first order, however, the gas-phase, i.e., \ion{H}{ii}-region averaged, metallicity should provide a fair representation of the chemical evolution of the galaxy as a whole. 

The galaxy hosting GRB~080605 has indeed a disturbed morphology, indicative of an early merger or intrinsically clumpy structure. A merger could have also triggered the enhanced star formation of the host of GRB~080605 when compared to GRB hosts at low redshift \citep[e.g.,][]{2009ApJ...691..182S, 2010AJ....139..694L}. GRB hosts with similar SFR, however, might not be uncommon at $z > 1$. A good fraction of X-ray selected GRB hosts at $1 < z < 2$ has observed $R$-band brightnesses (probing the rest frame UV) in a range between $24$~mag and $22.5$~mag \citep{2009AIPC.1111..513M} indicating dust un-corrected SFRs up to 10~$M_{\sun}\rm{yr}^{-1}$. Already mild dust-attenuation in the host can easily increase this to values of 50~$M_{\sun}\rm{yr}^{-1}$ or even higher, illustrating that GRB hosts with SFRs significantly above 10~$M_{\sun}\rm{yr}^{-1}$ are not an exceptionally rare phenomenon at $z > 1$ \citep[see also e.g.,][]{2005ApJ...633..317F, 2011ApJ...727L..53C, 2011arXiv1108.0674K, 2011arXiv1110.4642S}. %An illustrative low-$z$ example is the host of GRB~051022 \citep{2007A&A...475..101C}

\subsection{Afterglow versus host properties}

The substantial gas-phase metallicity of $Z\sim Z_{\sun}/2$ might directly relate to the substantial $A_V \sim 0.5$~mag including the presence of the 2175~\AA~dust feature as observed in the afterglow SED \citep{2011A&A...526A..30G} and spectrum \citep{2011Zafar}. A metallicity of around solar was also inferred from GRB-DLAs for GRBs~070802 and 080607, both of which were substantially reddened, and had 2175~\AA~dust features \citep{2009ApJ...697.1725E, 2009ApJ...691L..27P, 2010arXiv1009.0004P} as well. This seems to support the association between the 2175~\AA~bump and chemically evolved galaxies \citep[e.g.,][]{2009A&A...499...69N}. With only a small handful of such events, however, no strong conclusions can be drawn, yet.

\subsection{The mass-metallicity relation at $z\sim 2$}

Having the key parameters of stellar mass, metallicity and SFR of the host of GRB~080605 at hand, we can now investigate its relation to the mass-metallicity ($M_*$-$Z$) relation at $z\sim2$ \citep[e.g.,][]{2006ApJ...644..813E}. A further basic property is the host's location with respect to the fundamental metallicity relation (FMR) defined by SDSS galaxies in a mass range between $ 9.2 \lesssim  \log (M_*) \lesssim 11.4$. The FMR connects $M_*$, $12+\log(\rm{O}/\rm{H})$, and SFR \citep{2010MNRAS.408.2115M} via:

\begin{equation}
\label{fmp}
12+\log(\rm{O}/\rm{H}) = 8.90 + 0.47\times(\mu_{0.32}-10)
\end{equation}
where $\mu_{0.32} = \log (M_*\,[M_{\sun}]) - 0.32 \times \log($SFR$_{\rm H\alpha}\,[M_{\sun}\,\rm{yr}^{-1}])$. {The oxygen abundance for GRB~080605 on the \citet{2010MNRAS.408.2115M, 2011MNRAS.414.1263M} scale is $12+\log(\rm{O}/\rm{H}) = 8.52\pm0.09$. The value derived from $M_*$ and SFR via Eq.~\ref{fmp} is consistent with it ($12+\log(\rm{O}/\rm{H}) = 8.63\pm 0.08$). Errors are again based on the statistical uncertainty of line-flux measurement, correction factor and stellar mass only\footnote{A systematic error on the stellar mass estimate of $\pm0.2$~dex., for example, would translate into additional systematic errors of $\pm 0.12$ on the derived metallicity.}.}

This establishes the host of GRB~080605 as a star-forming galaxy which has no significant deficit of metals with respect to star-forming galaxies at low redshift for its given mass and SFR. Or, conversely, the selection through the energetic GRB~080605 does not lead to its host being metal-poor with respect to field galaxies of comparable stellar mass and SFR.

\begin{figure}
\centering
\includegraphics[width=\columnwidth]{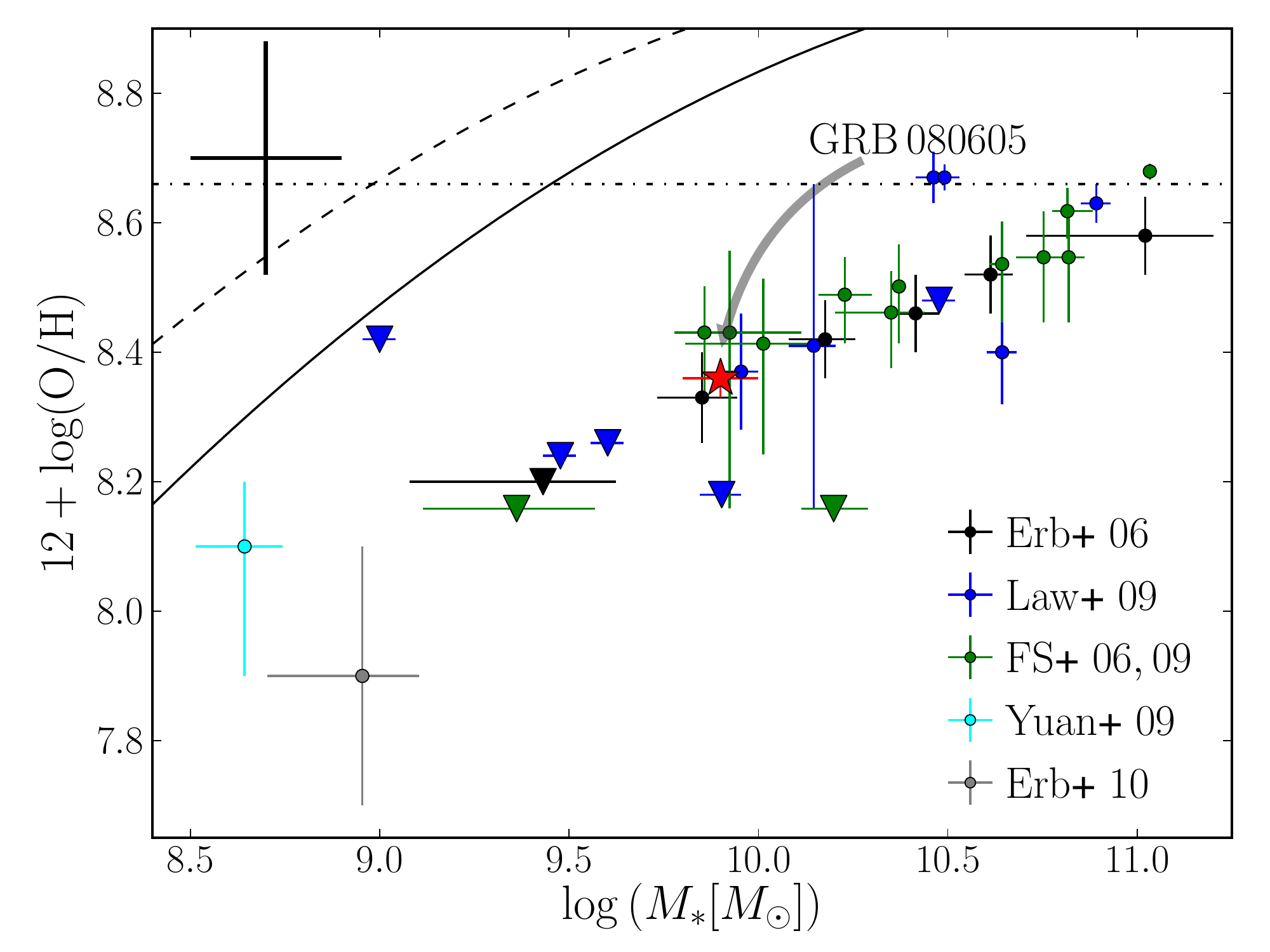}
\caption{The host of GRB 080605 with respect to the mass-metallicity relation at $z\sim2$. Different colored symbols represent the averaged galaxy distribution from \citet{2006ApJ...644..813E} in black, as well as individual sources from \citet{2006ApJ...645.1062F, 2009ApJ...706.1364F} in green, from \citet{2009ApJ...697.2057L} in blue and gravitationally lensed galaxies from \citet{2009ApJ...699L.161Y} and \citet{2010ApJ...719.1168E} in cyan and grey, respectively. Upper limits are shown with downward triangles with the same color-coding. All measurements are in the N2 scale of \citet{2004MNRAS.348L..59P}. The horizontal dashed-dotted line marks the solar oxygen abundance. {The dashed line is the local $M-Z$ relation \citep{2004ApJ...613..898T}, which is also shown shifted (solid line) to the observations at $z\sim 0.7$ \citep{2005ApJ...635..260S}}. Approximate systematic errors on the N2 metallicity scale and the mass determination are indicated in the top left corner.}
\label{fig:mmetz}
\end{figure}

The host of GRB~080605 hence provides the opportunity to probe the mass-metallicity relation at $z\sim 2$ \citep[e.g.,][]{2006ApJ...644..813E} at lower stellar masses (Fig.~\ref{fig:mmetz}). If populated with more events, GRB hosts can thus provide unique constraints on the low-mass end of the $M_*-Z$ relation \citep[see also e.g.,][]{2011arXiv1107.3841V} similar to measurements via gravitationally lensed objects \citep[e.g.,][]{2012arXiv1202.5267W} but without the need for (and uncertainty of) a detailed lens model (Fig.~\ref{fig:mmetz}).

\subsection{The non-detection of Ly$\alpha$}
\label{LyA}

The luminosity-independent selection of star-forming galaxies through GRBs offers a unique probe of the escape fraction ($f_{\rm{esc}}$) of Ly$\alpha$ photons. The path length of resonantly scattered  Ly$\alpha$ photons depends on the geometry and kinematics of \ion{H}{i} within a galaxy, and could thus be greatly enhanced as compared to, for example, the path length of photons from  recombination lines such as H$\alpha$. The longer path length directly translates into a higher dust absorption probability for Ly$\alpha$ photons  and hence $f_{\rm{esc}}$ might end up anywhere below unity \citep[e.g.,][]{2009A&A...506L...1A}. 

Ly$\alpha$ emission from GRB hosts was detected in both narrow-band imaging and afterglow/host spectroscopy \citep[e.g.,][]{2003A&A...406L..63F, 2005MNRAS.362..245J, 2010A&A...522A..20D, Bo2012}. The broad wavelength coverage of X-shooter extending down to the UV (Ly$\alpha$ line at $z\sim 1.64$ is redshifted to 3210~\AA) coupled with the tight constraints on the galaxies reddening and extreme luminosity of H$\alpha$, makes the host of GRB~080605 an ideal test case for the escape fraction in a high-redshift environment. 

At $f_{\rm{esc}} = 1$, the intrinsic ratio between Ly$\alpha$ and H$\alpha$ is 8.7 \citep{1971MNRAS.153..471B}. Consequently, Ly$\alpha$ is expected to be a factor $12$ more luminous than our non-detection implies. {This corresponds to an escape fraction of $f_{\rm{esc}} < 0.08$,  which was estimated in the same way as the flux limit but using the photometry-matched spectrum and its errors as discussed in Section~\ref{Xsred}.} %Using the most conservative assumptions, the escape fraction is still constrained to be $f_{\rm esc} < 0.14$. 
%This estimate does depend somewhat on the assumption of the extinction-law, as the steepness of its UV-slope, and thus the dust-correction factor for Ly-$\alpha$ differs significantly for various extinction laws.

While the evidence for reddening from the recombination lines and the stellar continuum is weak, the properties of the afterglow \citep{2011Zafar} provide compelling evidence that there is enough dust in the ISM to absorb the scattered Ly$\alpha$ photons efficiently. 

Our limit is consistent with previous estimates using narrow-band surveys targeting both Ly$\alpha$ and H$\alpha$ \citep{2010Natur.464..562H} or measured from the column density distribution of GRB-DLAs \citep{2009ApJS..185..526F}. A larger sample of hosts observed in similar fashion can provide competitive constraints on the average escape fraction in high-redshift environments at $1.6 < z < 2.5$. These measurements would be completely independent on conventional selection techniques, and representative of young, star-forming galaxies common in the early Universe. Establishing the average escape fraction at cosmological distances and for typical star-forming galaxies has strong implications for the use of Ly$\alpha$ emission as a tracer of star-formation and luminosity functions derived from Ly$\alpha$ galaxies at the highest redshifts. 

\section{Conclusions}

{We presented medium-resolution optical/NIR spectroscopy and ground and space-based imaging of the galaxy selected through GRB~080605 at $z=1.64$. Our HST imaging probes and resolves the large-scale structure of the host, and shows it to be a morphologically complex system that consists of two components separated by 8.6~kpc.} An X-shooter spectrum covering its rest-frame UV-to optical wavelength range (1150 to 8700~\AA) reveals a wealth of emission lines, including \oii, \oiii, H$\beta$ as well as \nii~and H$\alpha$. These recombination and forbidden lines allow us to put unique constraints on the conditions of the ISM in the host. It is in particular the first robust measurement of the gas-phase metallicity of a GRB host at $z > 1$ using strong-line indicators based on \nii~($\lambda$6584). %This was made possible through the NIR capabilities and exquisite sensitivity of the novel X-shooter spectrograph at the VLT. 

The host of GRB~080605 is significantly enriched with metals with an oxygen abundance $12 + \log(\rm{O}/\rm{H})$ between 8.3 and 8.6 ($0.4\,Z_{\sun} < Z < 0.8\,Z_{\sun}$) for several different strong-line diagnostics. In addition, its stellar mass is $M_* = 8.0^{+1.3}_{-1.6} \times 10^9 M_{\sun}$ and the galaxy is extremely star-forming (SFR$_{\rm{H}\alpha} = 31^{+12}_{-6}\,M_{\sun}\,\rm{yr}^{-1}$, sSFR$_{\rm{H}\alpha} = 4$~Gyr$^{-1}$). With a gas-phase metallicity above 40\% of the solar value and luminosity above $L^{*}$ \citep{2011arXiv1108.0674K}, it contrasts many observation of GRBs at lower redshift, which typically showed their hosts to be sub-luminous and metal-poor galaxies. Coupled with the high energy-release in $\gamma$-rays of $E_{\gamma, \rm iso} \sim 2.2\times10^{53}$~erg, it challenges those GRB progenitor models in which the formation of energetic GRBs requires very low metallicities.

{The metallicity measurement of the host of GRB 080605 directly shows that GRB hosts at $z > 1$ are not necessarily metal-poor, both on absolute scales as well as relative to their stellar mass and SFR.} Our detailed spectroscopic observations in fact suggest that the hosts of GRBs in general might provide a fair representation of the high-redshift, SFR-weighted population of ordinary star-forming galaxies. 

GRB hosts thus offer a selection of star-forming galaxies at high redshifts, including objects in the low-mass ($M_* \lesssim 10^{10} \,M_{\sun}$) regime, which are challenging to study otherwise. Targeted spectroscopic investigation become feasible through the afterglow's redshift, its sub-arcsec position and the substantial star-formation within GRB-selected galaxies. %GRBs thus provide a unique selection criteria for efficient studies of typical star-forming galaxies, which are challenging to access by other means.

Similar data for a representative and statistically significant sample of GRB hosts hold the key for understanding the nature of GRB hosts in particular and give important insights into the high-redshift population of star-forming galaxies in general. Furthermore, they yield the fundamental information to establish GRBs as probes of the star-formation up to the era of re-ionization. With the availability of highly redshift-complete GRB, afterglow and host samples such as TOUGH\footnote{\texttt{http://www.dark-cosmology.dk/TOUGH}} \citep[][Malesani et al., in prep.]{Jens2012, Palli2012, Bo2012, Tom2012, Michal2012} and NIR spectroscopy with X-shooter these studies are now feasible for the first time, and will continue to open the window with respect to the properties of GRB hosts in the previously unexplored redshift range $1 \lesssim z \lesssim 3$.

\begin{acknowledgements}
We thank L. Christensen and S. Savaglio for important insights and valuable discussion, and A. Rau for providing data for Figure 5 in machine-readable format. We also thank the referee for constructive comments, that helped to improve the quality of the manuscript. TK acknowledges support by the European Commission under the Marie Curie Intra-European Fellowship Programme in FP7. JPUF acknowledges support from the ERC-StG grant EGGS-278202. The Dark Cosmology Centre is funded by the Danish National Research Foundation. Based on observations made with the NASA/ESA Hubble Space Telescope, obtained from the data archive at the Space Telescope Institute.
\end{acknowledgements}

%\bibliography{./bibtex/refs,./bibtex/mnemonic}
%\bibliographystyle{./bibtex/aa}

\hyphenation{Post-Script Sprin-ger}

\end{document}